\documentclass[12pt]{article}

\usepackage{amsfonts}
\usepackage{amsmath}
\usepackage{esint}
\usepackage[margin=1in]{geometry}
\usepackage{graphicx}
\usepackage{epstopdf}
\usepackage{amssymb}
\usepackage{cite}
\usepackage{multicol}
\usepackage{caption}
\usepackage{color}

\usepackage{hyperref}
\usepackage{upgreek}
\usepackage{textcomp}
\usepackage{dsfont}
\usepackage{dblfloatfix}
\usepackage[dvipsnames]{xcolor}

\title{Identification of Tissue Optical Properties During Thermal Laser-Tissue Interactions: An Ensemble Kalman Filter-Based Approach}
\author{Andrea Arnold$^{1*}$ and~Loris Fichera$^2$}
\date{}

\begin{document}
\maketitle

\small 
{$1$ Department of Mathematical Sciences, Worcester Polytechnic Institute, Worcester, MA, USA}

{$2$ Department of Robotics Engineering, Worcester Polytechnic Institute, Worcester, MA, USA}

{$*$ Corresponding author}

\vspace{.5cm}

\centerline{E-mail: anarnold@wpi.edu, lfichera@wpi.edu}

\normalsize

\bigskip

\begin{abstract}
In this paper, we propose a computational framework
to estimate the physical properties that 
govern the thermal response of laser-irradiated tissue.
We focus in particular on two quantities, the \textit{absorption} and \textit{scattering}
coefficients, which describe the optical 
absorption of light in the tissue and whose knowledge
is vital to correctly plan medical laser
treatments.
To perform the estimation, we utilize an implementation of the
Ensemble Kalman Filter (EnKF), a type of Bayesian filtering
algorithm for data assimilation.
Unlike prior approaches, in this work we estimate the tissue optical
properties based on observations of the tissue thermal response to laser irradiation.
This method has the potential for straightforward implementation
in a clinical setup, as it would only require
a simple thermal sensor, e.g., a miniaturized
infrared camera.
Because the optical properties of tissue can undergo
shifts during laser exposure, we employ a variant of
EnKF capable of tracking time-varying parameters.
Through simulated experimental studies, we demonstrate the ability
of the proposed technique to identify the tissue optical
properties and track their dynamic changes
during laser exposure,
while simultaneously tracking changes in the tissue temperature at locations beneath the surface.
We further demonstrate the framework's capability in estimating additional unknown tissue properties (i.e., the volumetric heat capacity and thermal conductivity) along with the optical properties of interest.\\

\end{abstract}


\section{Introduction}
\label{sec:introduction}

Lasers are an integral part of modern
medicine, and their applications span 
across a wide range of therapeutic
areas~\cite{Niemz2019}.
In minimally invasive surgery, lasers are
frequently used to perform precise
tissue cutting and ablation~\cite{Fichera2021}.
Another major application area is photothermal therapy, 
where lasers are used to thermally necrotize
diseased tissue \textit{in-situ}, e.g., to treat
otherwise inoperable tumors~\cite{Thomsen2010}.
In all of these applications, it is of vital importance to control the interactions that occur between
the laser light and the tissue being treated.
Prior research has extensively explored the
mechanisms of light absorption in biological tissue,
and it has produced models capable of predicting
the physical tissue changes created by laser 
light~\cite{Niemz2019,Welch2011}.
These models can be used for treatment planning,
i.e., to determine what laser settings should be 
used to achieve the desired clinical outcomes.
Typically available settings include the laser wavelength,
beam waist, number and duration of pulses, 
and energy per pulse.

Existing laser-tissue interaction
models require explicit
knowledge of the tissue optical characteristics,
including the \textit{absorption} and
\textit{scattering} coefficients.
Taken together, these two coefficients describe the optical
penetration of light into the tissue and determine
what fraction of light is absorbed.
Knowledge of these coefficients is necessary to 
correctly plan and carry out a laser treatment,
but unfortunately accurate estimations
may not always be readily available.
Prior studies have experimentally 
documented absorption and scattering
in a variety of different biological
media~\cite{Kim2010,Jacques2013}.
In practice, the applicability of these results is
limited by the fact that the optical properties of
living tissue can vary considerably from
individual to individual, site to site,
and even time to time~\cite{Jacques2013}.
It would be desirable to measure the
tissue optical properties
directly during a procedure, so that the
operating physician could regulate the laser
settings accordingly.

Motivated by the foregoing considerations, 
in this paper we propose a new method to
determine the absorption and scattering
coefficients of living tissue during a laser
treatment.
Because absorption and scattering can be difficult
to characterize via direct measurements, most
of the techniques currently available 
for this purpose are based on indirect
estimation methods.
A common approach is to measure the tissue
reflectance, either with an integrating
sphere~\cite{Foschum2020,Cook2020} or
a specialized fiberoptic
device~\cite{Yu2014,Gunther2020,Tanis2016}, and
then inversely fit the unknown absorption and
scattering coefficients using a model of light
propagation in tissue.
These methods are effective, but they may require the
introduction of bulky and/or expensive
instrumentation in the clinical setup.
In contrast to existing techniques, in this paper
we propose to characterize absorption and
scattering based on the observation of the
tissue thermal response to laser exposure.
The rationale for this approach is provided by
the fact that the tissue temperature is 
routinely monitored during many medical laser
procedures~\cite{Saccomandi2013}.
Therefore, the method we propose
in this paper would be relatively straightforward to
implement in the workflow of a laser procedure,
as in fact it would not require the
introduction of additional instrumentation
in the clinical setup.

To characterize absorption and scattering, we 
propose to use an Ensemble Kalman Filter (EnKF),
a type of computational Bayesian filtering algorithm for data assimilation~\cite{Evensen2009,Fearnhead2018,Arnold2014,Katzfuss2016}.
Briefly, our proposed approach works as follows:
given an initial guess of the unknown coefficients,
we first predict the tissue thermal response
using a thermal laser-tissue interaction model;
we then update the tissue optical
properties to minimize the difference between
the predicted and observed tissue temperature.
The sequential nature of EnKF algorithms enables the
implementation of an \textit{online} estimation process,
i.e., the absorption and scattering coefficients are 
progressively refined as more and more temperature
observations become available over time.
In particular, we utilize a version of the EnKF capable of estimating time-varying parameters~\cite{Arnold2019,Campbell2020,Arnold2020}.
As we show later in this paper, our approach enables the
detection and tracking of
dynamic changes in the tissue properties that may occur
during laser exposure~\cite{Vogt2018,Nagarajan2020}, while
simultaneously predicting the
temperature of the tissue at unmeasured locations
beneath the tissue surface.

\subsection{Contributions}

The main contributions of this paper are as follows:
\begin{itemize}
\item We propose a new computational framework which utilizes ensemble Kalman filtering and thermal sensor measurements from the surface of the tissue to identify the tissue optical properties during thermal laser-tissue interactions. Key features of the proposed approach include: (i) the ability to detect and track dynamic changes in the tissue properties during laser exposure; and (ii) the capability of simultaneously predicting the temperature of the tissue at unmeasured locations beneath the tissue surface. We provide a comprehensive description of the framework, which enables other researchers to replicate this work and integrate it in their own setups.
\item Through a set of numerical experiments with increasing complexity, we demonstrate the viability of the proposed approach in estimating both constant and time-varying tissue optical properties (namely, the absorption and scattering coefficients), while simultaneously tracking the temperature of the tissue at locations beneath the tissue surface, given sequential temperature measurements obtained at a single location on the tissue surface. We further demonstrate the ability of the proposed framework to estimate additional unknown tissue properties (i.e., the volumetric heat capacity and thermal conductivity) along with the optical properties of interest.
\end{itemize}
We note that while Kalman filtering is a well-known data assimilation technique, our work uses a formulation of the EnKF that allows for the online tracking of time-varying parameters \cite{Arnold2019,Arnold2020,Campbell2020}. Such capability is vital in the estimation of tissue optical properties, as prior research shows that these properties can shift during laser exposure \cite{Vogel2003,Vogt2018,Nagarajan2020}.
The approach outlined in this paper demonstrates a novel application of the EnKF in this setting with potential to advance the state-of-the-art for tissue identification during medical laser procedures.

\subsection{Paper Outline}

The remainder of the paper is organized as follows:
Section~\ref{sec:methods} presents the proposed method, first briefly reviewing
the dynamics of thermal laser-tissue interactions 
and then formulating the tissue identification problem;
Section~\ref{sec:experiments}
describes the simulations performed
to verify the viability of the proposed approach;
Section~\ref{sec:discussion}
discusses 
the contributions and limitations of this study;
finally,
Section~\ref{sec:conclusion} concludes the manuscript.


%
\section{Materials and Methods}
\label{sec:methods}

In this section, we begin with an overview of our proposed approach for identifying the tissue optical properties given thermal sensor measurements. We then briefly review the role that the absorption and scattering coefficients play
in the thermal response
of laser-irradiated tissue.
We provide an overview of
the dynamics of thermal laser-tissue interactions and
introduce the temperature model used 
by the EnKF.
A detailed derivation of this model is beyond the scope
of this paper, and interested readers are referred to
available textbooks on the topic; see, e.g., \cite{Niemz2019}.
Finally, we describe the procedure for estimating the absorption and scattering
coefficients using the EnKF with thermal sensor data.

\subsection{Online Estimation of Tissue Optical Properties}

Our proposed online estimation approach to identify the tissue absorption
and scattering coefficients is based on using ensemble Kalman filtering to assimilate thermal sensor data. 
In the EnKF framework, the unknown parameters of a system
are modeled as stochastic variables whose probability
density functions are represented by a set of random
realizations called an~\textit{ensemble}.
Each time new thermal sensor measurements become available, 
the ensemble is manipulated through a set of update
rules to reflect the new probability distributions of the 
unknown parameters conditioned on the observed data.
In particular, we utilize a version of the EnKF capable of tracking
time-varying parameters~\cite{Arnold2019,Campbell2020},
which enables us to monitor the shift
in the tissue optical properties
that may occur during a laser
procedure~\cite{Vogt2018,Nagarajan2020,Vogel2003}.

\begin{figure}[t!]
  \centerline{\includegraphics[width=0.6\linewidth]{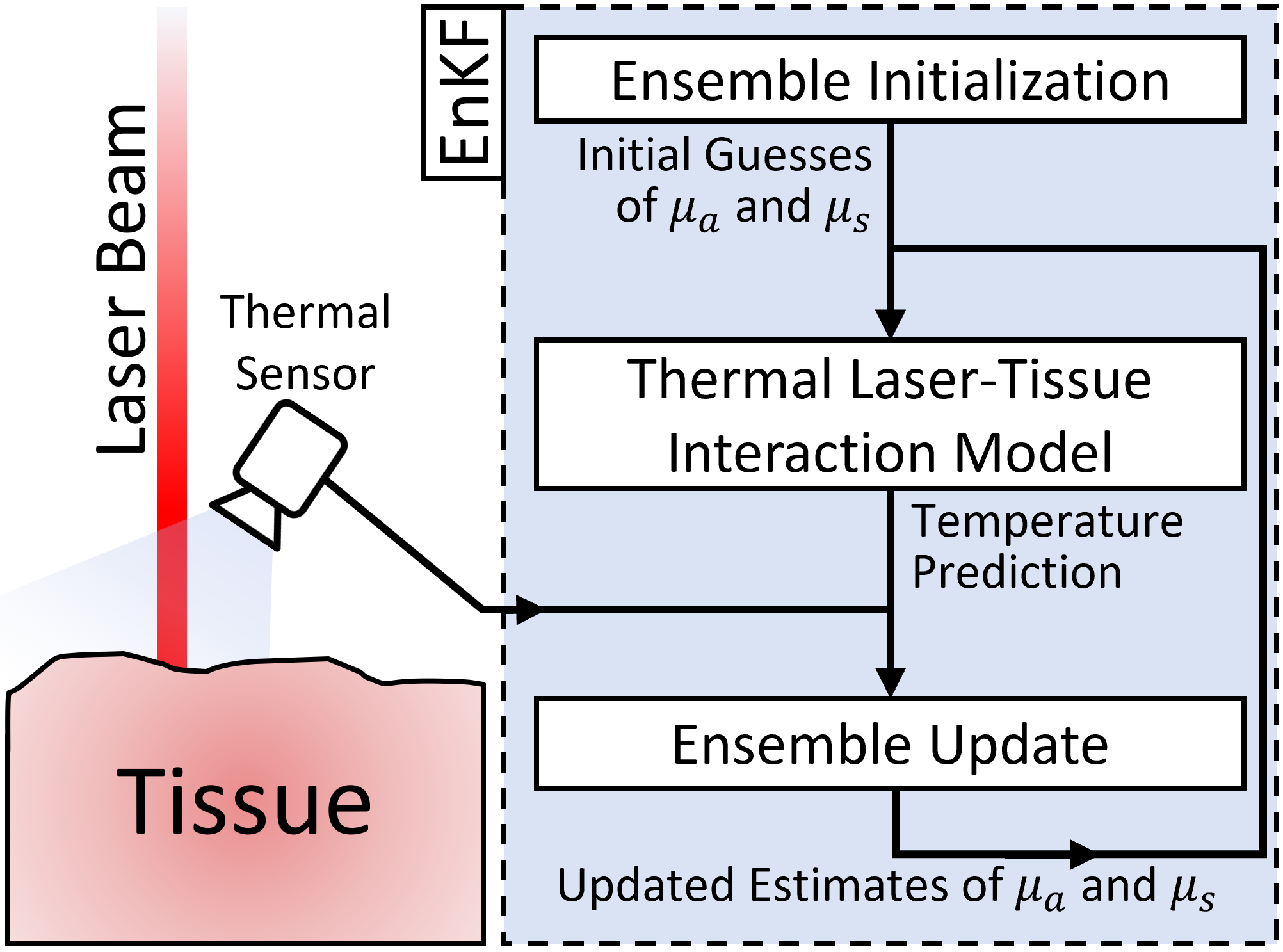}}
  \caption{Proposed approach for the identification of the 
  tissue absorption and scattering coefficients, $\mu_a$
  and $\mu_s$, respectively, during laser surgery.
  The laser light delivered to the tissue is absorbed
  under the form of heat, and the corresponding
  temperature increase is observed with a thermal sensor,
  i.e., an infrared thermal camera.
  An Ensemble Kalman Filter (EnKF) is used to estimate the 
  unknown coefficients based on the observed temperature
  dynamics.
  }
  \label{fig:approach}
\end{figure}

Figure~\ref{fig:approach} illustrates the 
proposed estimation approach: 
the tissue is irradiated with a laser pulse,
triggering a localized temperature increase which
is observed by a thermal sensor;
with each observation, the EnKF compares the sensor data with the output
of a laser-tissue interaction model and updates
the ensemble in such a way to minimize the error between
the model output and the measured tissue temperature.
At any given time, the ensemble mean is used as an
estimate of the unknown absorption and scattering
coefficients, while the standard deviation provides a
measure of uncertainty.

\subsection{Thermal Response of Laser-Irradiated Tissue}
\label{sec:forward-model}

To model the tissue's thermal response to laser irradiation, we
consider a scenario where a block
of tissue is exposed in air to a laser 
beam. 
We further assume that the laser beam is perpendicular to the tissue surface.
A Cartesian reference frame is established on the surface
of the tissue so that the $Z$-axis of the frame corresponds
to the optical axis of the laser beam.
We define the tissue temperature as a function $T(x,y,z,t)$,
where $x$, $y$, $z$ are spatial coordinates and $t$ 
represents time.
The tissue temperature can be calculated by solving the
following differential equation~\cite{Niemz2019}:
\begin{equation}
c_v \frac{\partial{T}}{\partial{t}} =
k \left(\frac{\partial^2{T}}{\partial{x}^2}
+ \frac{\partial^2{T}}{\partial{y}^2}
+ \frac{\partial^2{T}}{\partial{z}^2}
 \right) + S
 \label{eq:heat-equation}
\end{equation}
where $c_v$ is the volumetric heat
capacity of the tissue (J~cm$^{-3}$~K$^{-1}$), $k$ is the tissue
thermal conductivity (W cm$^{-1}$ K$^{-1}$),
and $S$ is the volumetric power density
(W cm$^{-3}$).
This latter term models the heat created by the
laser in the tissue and is given by
\begin{equation}
    S = PA
    \label{eq:volumetric-power-density}
\end{equation}
where $P$ is the beam power (W) and $A$ is the
light absorption map (cm$^{-3}$),
which represents the 
fraction of light captured at any given location 
within the tissue volume~\cite{Jacques2010}.

To calculate the absorption map $A$, it is necessary to 
model the diffusion of light into the tissue.
It is in
these calculations that the coefficients of absorption 
$\mu_a$ and scattering $\mu_s$ (both having units of cm$^{-1}$) 
appear.
Obtaining a closed form solution for the absorption map
can be challenging, and this quantity is frequently
calculated with a Monte Carlo 
method instead~\cite{Jacques2010,Marti2018}:
the idea is to simulate the optical path of a large number
of photons as a discrete random walk,
and to keep track of where the photons deposit energy.
The length of each step of the walk is sampled from 
a logarithmic distribution~\cite{Jacques2010}, i.e.,
\begin{equation}
    s = \frac{-\ln(\zeta)}{\mu_a+\mu_s}
    \label{eq:mc-step-size}
\end{equation}
where $\zeta$ is a computer-generated number sampled
uniformly at random between 0 and 1.
When a photon moves from one step to the next, its 
direction of travel will change due to scattering.
This change in direction is modeled by means of
an azymuthal component, sampled uniformly at random
between 0 and 2$\pi$, combined with a deflection 
angle $\alpha$, which is typically modeled using the Henyey-Greenstein function~\cite{Henyey1941}, i.e.,
\begin{equation}
  p(\cos(\alpha)) = \frac{1-g^2}{2(1+g^2-2g\cos(\alpha))^{3/2}}
\end{equation}
with $g$ being the expected value of $\cos(\alpha)$.
This parameter is also known as the \textit{anisotropy
factor}, and for most biological tissues, its value has
experimentally been determined to range between 0.7 and
0.99~\cite{Niemz2019}.

\begin{figure}[t!]
  \centerline{\includegraphics[width=\linewidth]{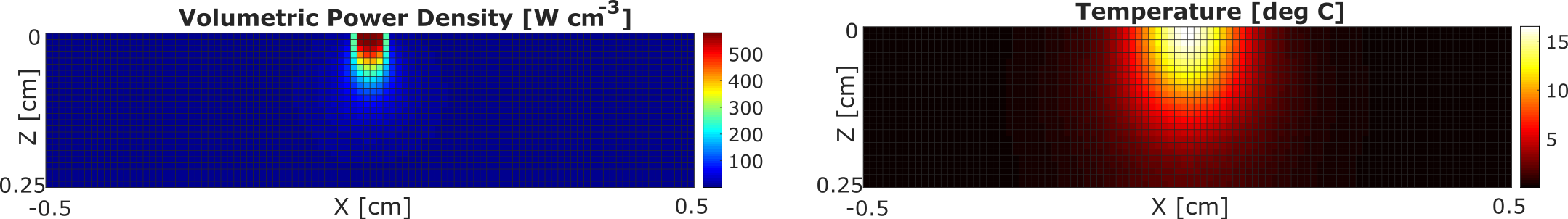}}
  \caption{
  (Left) Volumetric power density
  created by a Gaussian laser beam (waist: 250 $\upmu$m)
  in a block of tissue with dimensions
  1~cm~$\times$~1~cm~$\times$~0.25~cm and properties listed in
  Table~\ref{tab:tissue-parameters}.
  (Right) Tissue temperature 
  after a 0.5 s laser pulse, followed by a 0.5 s cooling
  phase.
  The initial tissue temperature was 0~\textdegree C.
  }
  \label{fig:simulation-example}
\end{figure}

At each step of the walk, a photon loses a fraction 
of its energy due to absorption.
A photon is terminated either when it escapes
the tissue volume or when its residual energy level falls
below some arbitrary small positive value. 
Figure~\ref{fig:simulation-example} shows an example of
the volumetric power density $S$ generated by a laser beam
and the temperature gradient that is created as a 
result in the tissue.

\subsection{Parameter Estimation via Ensemble Kalman Filtering}
\label{sec:param-estimation}

To estimate and track potential changes in the tissue optical properties, we utilize an augmented version of the EnKF for combined state and parameter estimation.
In this work, the tissue temperature $T$ at specified locations is considered as the state of the model, and the absorption and scattering coefficients, $\mu_a$ and
$\mu_s$ in \eqref{eq:mc-step-size}, are the parameters of interest.
Given the observed thermal sensor data, our goal is to formulate an approximation of the joint probability density function $\pi(T,\mu_a,\mu_s)$ using a discrete sample.
For conciseness of notation, we introduce a parameter vector
$\theta = (\mu_a,\mu_s)$, so that the probability
density function can simply be written as
$\pi(T,\theta)$.

Assume that we have a set of measurements $d_j$ of the tissue temperature obtained sequentially by the thermal sensor at discrete times $t_j$, with $j=1,\dots,M$.
Further assume that the data are corrupted by measurement errors. 
Let $T_j$ denote the temperature predicted by the 
laser-tissue interaction model 
described in Section~\ref{sec:forward-model}
at time $j$, and let
$\theta_j = (\mu_{a,j},\mu_{s,j})$ be a vector
containing the parameter estimates at time $j$.  
The filtering process begins by drawing a random sample
of size $N$ from the prior distribution
$\pi(T_0,\theta_0)$, which encodes any prior knowledge
on the unknown coefficients. This forms the initial ensemble at time $j=0$. The filter then proceeds in a two-step updating scheme from time $j$ to $j+1$, with the prediction and analysis steps detailed as follows.

\subsubsection{Prediction Step}
\label{sec:prediction-step}

Given the current ensemble $\mathcal{S}_j = \big\{ (T_j^n,\theta_j^n) \big\}_{n=1}^N$ at time $j$, the prediction step updates the temperature values using a model approximation; i.e.,
\begin{equation}
    T_{j+1\mid j}^n = F(T_j^n,\theta_j^n) + v_{j+1}^n, \quad n = 1,\dots, N
\end{equation}
where $F(T_j^n,\theta_j^n)$ represents the numerical solution to \eqref{eq:heat-equation} at time $j+1$, stored as a column vector, and the innovation $v_{j+1}^n$ accounts for uncertainty in the forward prediction. 
The innovation term is typically drawn from a Gaussian distribution with zero mean and some prescribed covariance; i.e., $v_{j+1}^n\sim\mathcal{N}(0,\mathsf{C})$, where $\mathsf{C}$ is a diagonal matrix set prior to running the filtering algorithm.
The parameter values $\theta_j^n$ are propagated forward using a random walk model of the form
\begin{equation}\label{Eq:RW_model}
    \theta_{j+1\mid j}^n = \theta_j^n + \xi_{j+1}^n, \quad n = 1,\dots, N
\end{equation}
where $\xi_{j+1}^n \sim\mathcal{N}(0,\mathsf{E})$ with a prescribed covariance matrix $\mathsf{E}$, 
which is also set prior to running the filter.
Note that the parameter forward prediction in \eqref{Eq:RW_model} is not necessary if the parameters are assumed to be constants; however, it is vital in tracking time-varying parameters \cite{Campbell2020}.
The predicted temperature and parameter values are augmented into vectors of the form
\begin{equation}
    z_{j+1\mid j}^n = \left[\begin{array}{c} T_{j+1\mid j}^n \\ 
    \theta_{j+1\mid j}^n
    \end{array}\right] , \quad n = 1,\dots,N
    \label{eq:ensemble}
\end{equation}
which are used to compute the sample mean
\begin{eqnarray}\label{Eq:Ens_Mean}
    \bar{z}_{j+1\mid j} = \frac{1}{N}\sum_{n=1}^N z_{j+1\mid j}^n
\end{eqnarray}
and covariance matrix
\begin{eqnarray}\label{Eq:Ens_Cov}
    \mathsf{\Gamma}_{j+1\mid j} = \frac{1}{N-1}\sum_{n=1}^N (z_{j+1\mid j}^n-\bar{z}_{j+1\mid j})(z_{j+1\mid j}^n-\bar{z}_{j+1\mid j})^\mathsf{T}
\end{eqnarray}
of the prediction ensemble. 
Note that the covariance matrix $\mathsf{\Gamma}_{j+1\mid j}$ contains the variances of the predicted states and parameters along its main diagonal, with covariance information between pairs encoded in the off-diagonal entries.

\subsubsection{Analysis Step}

During the analysis step, the observed data $d_{j+1}$ are assimilated in producing the posterior ensemble, which is computed by
\begin{equation}\label{Eq:Analysis_Step}
    z_{j+1}^n = z_{j+1\mid j}^n + \mathsf{K}_{j+1} \bigg( d_{j+1}^n-G(z_{j+1\mid j}^n) \bigg)
\end{equation}
for each $n = 1,\dots, N$.  Here
\begin{equation}\label{Eq:Obs_Ens}
    d_{j+1}^n = d_{j+1} + w_{j+1}^n, \quad n = 1,\dots, N
\end{equation}
generates an ensemble of fictitious measurements around the observed data $d_{j+1}$, with $w_{j+1}^n \sim \mathcal{N}(0,\mathsf{D})$ representing observation error for some prescribed covariance matrix $\mathsf{D}$; $G$ is the observation function, which maps the predicted states and parameters to corresponding model observations; and $\mathsf{K}_{j+1}$ is the Kalman gain matrix, which contains cross-correlation information between the predicted model states and parameters.  
In this work, the observation function $G$ in \eqref{Eq:Analysis_Step} is a linear mapping
\begin{equation}
    G(z_{j+1\mid j}^n) = \mathsf{P} z_{j+1\mid j}^n, \quad n = 1, \dots, N
\end{equation}
where the projection matrix $\mathsf{P}$ picks out
the tissue location at which the temperature is being
measured, and the Kalman gain is computed by
\begin{equation}
    \mathsf{K}_{j+1} = \mathsf{\Gamma}_{j+1\mid j} \mathsf{P}^\mathsf{T} \big(\mathsf{P} \mathsf{\Gamma}_{j+1\mid j} \mathsf{P}^\mathsf{T} + \mathsf{D} \big)^{-1}.
\end{equation}
Since the tissue optical properties are not observed (i.e., there are no sequential measurements available for these quantities), the parameters are updated in the analysis step only through their cross-correlation with the tissue temperature, which is encoded in the Kalman gain; 
see, e.g., \cite{Arnold2014} for more details.
Posterior ensemble statistics are computed similarly as in \eqref{Eq:Ens_Mean} and \eqref{Eq:Ens_Cov} using $z_{j+1}^n$.

This two-step iterative process repeats for $j<M$. Note that here we assume that temperature data are available at each time $j$; if measurements are not available at a subset of the filter time steps, the observation update can be neglected such that the prediction ensemble serves as the posterior at these steps.  
Further note that $d_{j+1}$ in \eqref{Eq:Obs_Ens} represents the observed temperature measurement at time $j+1$, which we assume is available through a suitable thermal sensor (see Figure~\ref{fig:approach}). The action in \eqref{Eq:Obs_Ens} thereby accounts for uncertainty in this measurement by generating an ensemble of fictitious observations drawn from a Gaussian distribution whose mean is given by the observed data. This avoids under-estimating the covariance of the ensemble; see \cite{Burgers1998} for more details.


%
\section{Results}
\label{sec:experiments}

In this section, we present a set of numerical experiments demonstrating the viability of the proposed method in estimating the unknown tissue absorption and scattering
coefficients given simulated thermal sensor data. Simulations were performed using the MATLAB\textsuperscript{\tiny\textregistered} programming language (The MathWorks, Inc., Natick, MA). In the results that follow, we solved the thermal laser-tissue interaction model described in Section~\ref{sec:forward-model} using the toolbox developed by~\cite{Marti2018}. In this toolbox, the volumetric power density $S$ in \eqref{eq:volumetric-power-density} is calculated using a Monte Carlo method; then, the heat equation in \eqref{eq:heat-equation} is solved with a finite element method~\cite{Marti2018}.

\begin{figure}[t!]
  \centerline{\includegraphics[width=0.5\linewidth]{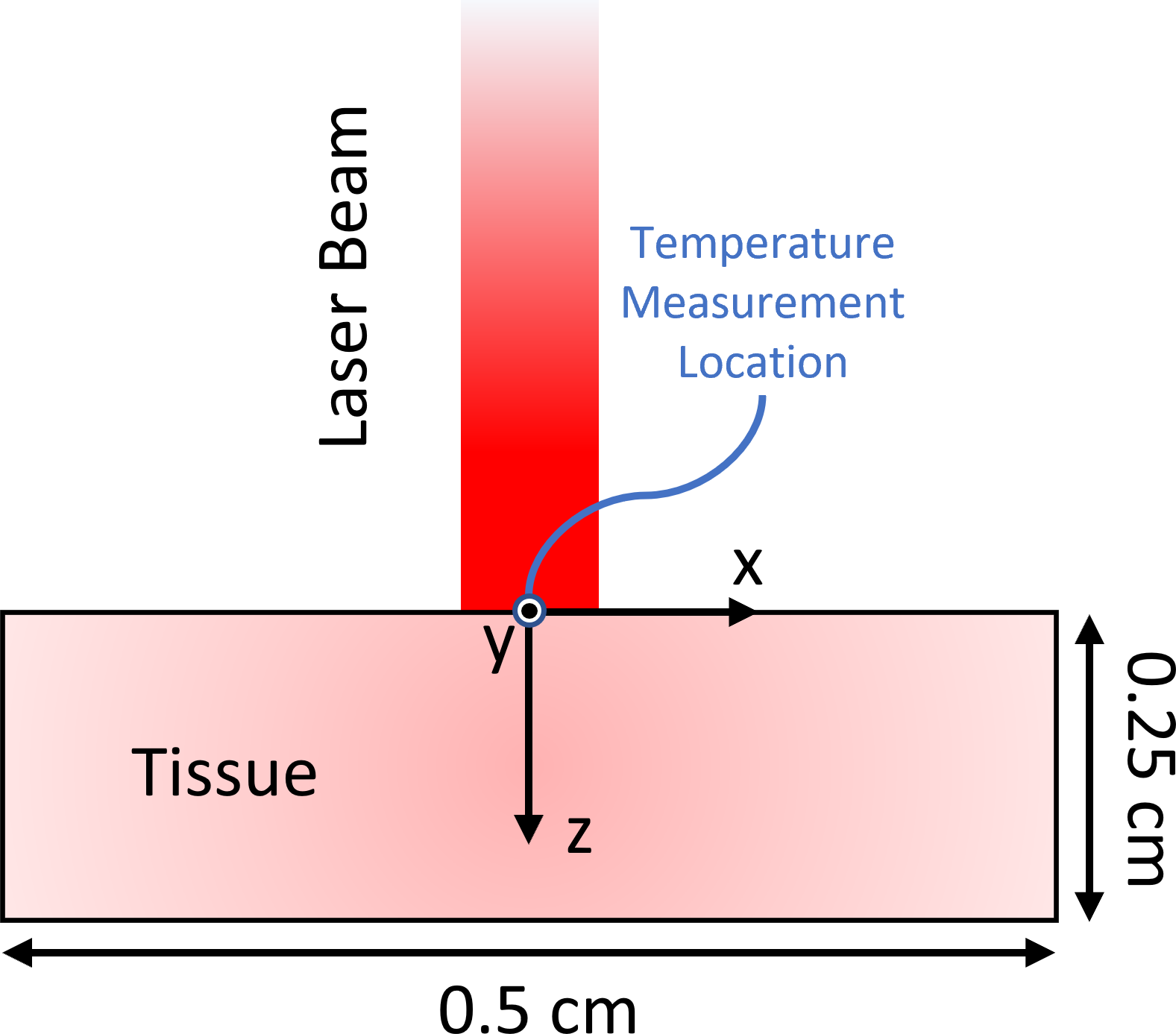}}
  \caption{Laser-tissue interaction setup reproduced in the simulation studies. A laser pulse
  is applied on the surface of a rectangular block of tissue, with dimensions as noted in the figure. A right-handed Cartesian reference frame is established on the surface of the tissue, with the $Z$-axis of the frame coincident with the optical axis of the laser beam. The $Y$-axis points out of the page.  Throughout the experiment, the tissue temperature is monitored at the origin of the reference frame. In a realistic setup, temperature measurement at this location would be possible with a non-contact thermal sensor, e.g., the one described in~\cite{Lin2017}. }
  \label{fig:simulation-setup}
\end{figure}


\subsection{Data Generation and Setup of Numerical Experiments}

In each experiment,
we simulate the setup 
illustrated in Figure~\ref{fig:simulation-setup},
where a laser pulse is applied to the
surface of a sample of biological tissue. 
As a baseline, we assume that the tissue has the physical
properties summarized in Table~\ref{tab:tissue-parameters}; 
these properties are similar to those used in \cite{Marti2018} to represent standard tissue.
The laser beam has a uniform intensity profile,
with a radius of 1 mm and a power of 0.5 W.
The dimensions of the tissue sample are
0.5 cm $\times$ 0.5 cm $\times$ 0.25 cm.
For computational purposes, the tissue block is
discretized into a grid of 100
$\times$ 100 $\times$ 50 cubic elements,
or \textit{voxels}.

We assume that tissue temperature measurements are available through a suitable sensor. Different thermal sensors have been proposed to monitor the temperature of tissue during laser irradiation. Non-contact infrared (IR) sensors provide a convenient option, as they can monitor the superficial tissue temperature without interfering with the laser application; see \cite{Lin2017,Pardo2014}. 
In this paper, we simulate the use of an IR sensor that monitors the temperature at the origin of the reference frame shown in Figure~\ref{fig:simulation-setup}.
More specifically, in the simulations that follow, the tissue
temperature at $(x,y,z) = (0,0,0)$ is monitored
with a virtual non-contact thermal sensor
which provides measurements at a rate of 10 Hz.
To simulate the presence of sensor noise, temperature
measurements at each time are altered with the addition of a Gaussian error term with zero mean and variance 0.01.

\begin{table}
\centering
\caption{Tissue physical properties used in the simulation studies.}
\label{tab:tissue-parameters}
\vspace{0.2cm}
\begin{tabular}{cccc}
\hline
\noalign{\vspace{0.1cm}}
\textbf{Symbol} & \textbf{Physical Variable} & \textbf{Units} & \textbf{Value}\\ 
\hline
\noalign{\vspace{0.1cm}}
$\mu_a$ & Absorption Coefficient & cm$^{-1}$ & 1\\[0.5mm]
$\mu_s$ & Scattering Coefficient & cm$^{-1}$ & 100\\[0.5mm]
$c_v$   & Volumetric Heat Capacity & J~cm$^{-3}$~K$^{-1}$ & 3.76\\[0.5mm]
$k$ & Thermal Conductivity & W cm$^{-1}$ K$^{-1}$ & 0.0037\\[0.5mm]
\hline
\end{tabular}
\end{table}

\begin{figure}[t!]
 \centerline{\includegraphics[width=\linewidth]{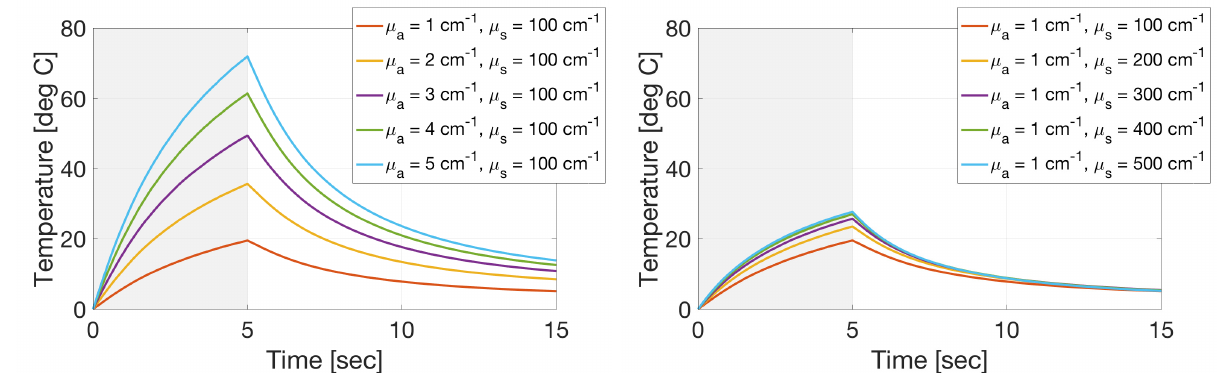}}
 \caption{Sensitivity of the simulated tissue temperature at the measurement location shown in Figure~\ref{fig:simulation-setup} with respect to the absorption (left) and scattering (right) coefficients. The volumetric heat capacity and thermal conductivity are fixed to the values in Table~\ref{tab:tissue-parameters}. In each plot, the laser pulse duration is shaded in gray.}
 \label{fig:sens}
\end{figure}

As noted, our goal is to identify the optical properties (i.e., the absorption and scattering
coefficients) of the tissue given the thermal sensor data. 
Before proceeding with the estimation, we first analyze the sensitivity of the simulated tissue temperature at the sensor location with respect to variations in the absorption and scattering coefficients. Figure~\ref{fig:sens} shows that the temperature at this location is more sensitive to relatively small changes in the absorption coefficient than the scattering coefficient: distinct temperature profiles are clear for each of the different values of $\mu_a$ (here ranging from 1 to 5 cm$^{-1}$), with each unit increase in $\mu_a$ corresponding to a significant increase in the peak temperature occurring at the end of the pulse duration; however, the temperature profiles are more difficult to distinguish for the different values of $\mu_s$ (ranging from 100 to 500 cm$^{-1}$), with increases in $\mu_s$ by increments of 100 cm$^{-1}$ resulting in comparatively small changes in the tissue temperature. While not shown, this distinction between temperature profiles becomes even more challenging in the presence of sensor noise.

In the numerical experiments that follow, we consider two cases: one in which the optical properties remain constant (static) throughout the laser-tissue interaction (as may occur during tissue probing); and one in which the optical properties shift during laser exposure (as may occur during a laser procedure).
In the former case, we assume that the tissue has the constant physical properties given in
Table~\ref{tab:tissue-parameters}; in the latter case, we modify the absorption and scattering coefficients to include time-varying dynamics.
While our focus in this work is on identifying the optical properties, the solution to the heat equation~\eqref{eq:heat-equation} also relies on knowledge of two additional tissue parameters, namely, the volumetric heat capacity, $c_v$, and thermal conductivity, $k$.
We note that these additional parameters may not always be
known \textit{a priori} in a realistic setting,
but reasonable approximations can generally be obtained using
empirical models available in the laser-tissue
interactions literature~\cite{Niemz2019}.
When these parameters are also unknown or uncertain, the proposed framework can be extended to accommodate their estimation; we demonstrate this in the results that follow.

At the beginning of each numerical experiment, we initialize the EnKF with an ensemble of size $N = 50$. 
We assume the initial tissue temperature is 0~\textdegree C and draw the initial values for the unknown absorption and scattering coefficients from uniform prior distributions, where the bounds are taken to be 0.5 to 2 times the values listed in Table~\ref{tab:tissue-parameters}.
To perform the temperature prediction step
(see Section~\ref{sec:prediction-step}), the filter runs its own implementation of the laser-tissue interaction model with a coarser tissue grid (i.e., 20 $\times$ 20 $\times$
10 voxels) than the one used to simulate the sensor data.
We equip the EnKF with a coarser geometric
tissue model for two reasons: (i) to
limit the computational complexity of the filter; and
(ii) to verify that the filter is able to
perform the estimation when a perfect geometric
model of the tissue is not available (as would be
the case in realistic application scenarios).
In addition to estimating the unknown tissue properties in each case, we also demonstrate the filter's ability to simultaneously track the tissue temperature at unobserved locations beneath the tissue surface. More specifically, we estimate the temperature profiles at the sensor location $(x,y,z)=(0, 0, 0)$ and at depths of 0.1 cm and 0.2 cm directly below, with respect to the tissue geometry shown in Figure~\ref{fig:simulation-setup}.


\subsection{Identifying Constant Absorption and Scattering}
\begin{figure}[t!]
  \centerline{\includegraphics[width=\linewidth]{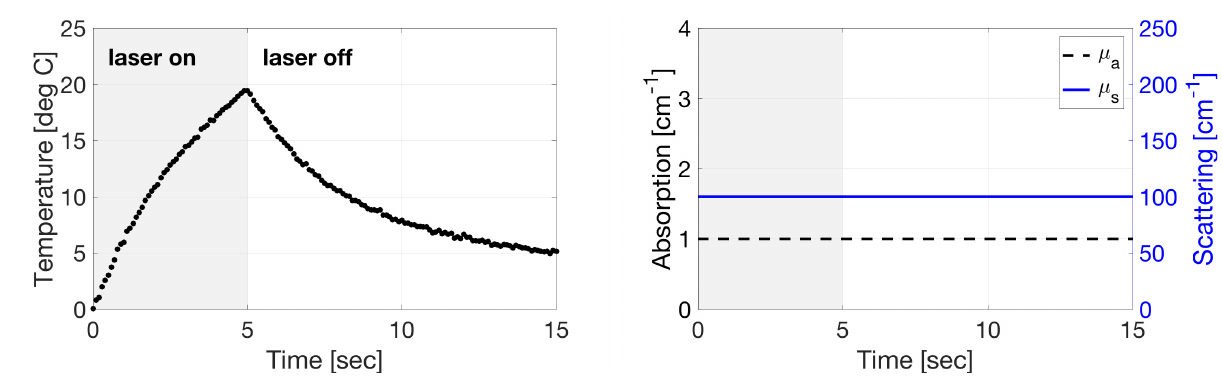}}
  \caption{Sensor data generated with constant tissue optical properties. (Left) Surface temperature data observed via virtual sensor at the measurement location shown in Figure~\ref{fig:simulation-setup}. (Right) The corresponding absorption and scattering coefficients, as given in Table~\ref{tab:tissue-parameters}. The volumetric heat capacity and thermal conductivity are also as given in Table~\ref{tab:tissue-parameters}. In each plot, the laser pulse duration is shaded in gray.}
  \label{fig:data_const}
\end{figure}

In the first experiment, we simulate a scenario in which the tissue optical properties remain constant during laser irradiation. 
This scenario may occur, e.g., in probing the tissue with a short low-power laser pulse to identify its optical properties and inform the planning of laser actions before a surgical procedure.
In this simulation, we apply a laser pulse for 5 seconds, then
continue to observe the tissue temperature for 10 more seconds.
Figure~\ref{fig:data_const} shows the simulated sensor observations and the corresponding constant absorption and scattering coefficients, as in Table~\ref{tab:tissue-parameters}.
Given the sensor data, and assuming knowledge of the other tissue properties, we aim to estimate the true values of the absorption and scattering coefficients.

\begin{figure}[t!]
  \centerline{\includegraphics[width=\linewidth]{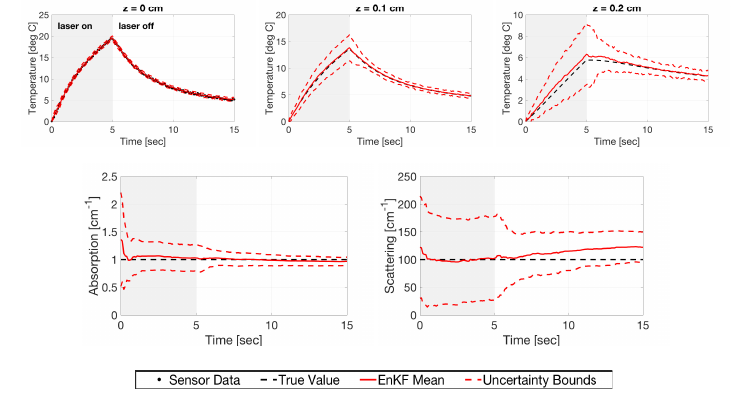}}
  \caption{Constant parameter estimation results. (Top) EnKF tissue temperature estimates at three locations on and below the surface of the tissue. (Bottom) EnKF estimates of the absorption and scattering coefficients. In each plot, the laser pulse duration is shaded in gray, the true value of the temperature or parameter is shown in dashed black, the EnKF mean is shown in solid red, and the $\pm 2$ standard deviation uncertainty bounds around the mean are shown in dashed red. The sensor data are shown in black markers on the temperature plot when $z=0$, overlaid with the EnKF estimate and uncertainty bounds.}
  \label{fig:results_constant}
\end{figure}

Figure~\ref{fig:results_constant} shows the resulting EnKF estimates of $\mu_a$ and $\mu_s$, along with the corresponding tissue temperature estimates at the three specified locations on and beneath the surface of the tissue; i.e., at the sensor location $(x,y,z)=(0, 0, 0)$ and at depths of $z = 0.1$ cm and $z = 0.2$ cm with respect to the tissue geometry in Figure~\ref{fig:simulation-setup}. Note that the EnKF estimate in each plot is the ensemble mean, with uncertainty bounds given by $\pm 2$ standard deviations around the mean.

As seen in Figure~\ref{fig:results_constant}, the EnKF provides an accurate estimate of the absorption coefficient, with uncertainty bounds shrinking over time. The estimate of the scattering coefficient drifts to a slightly higher value after the laser is turned off, with wider uncertainty bounds that do however contain the true parameter value. 
This drift may result from the fact that the thermal response of the tissue in \eqref{eq:heat-equation} is less sensitive to variations in scattering than absorption, as illustrated in Figure~\ref{fig:sens}. In fact, small variations in the scattering coefficient (here $<$ 25 cm$^{-1}$ between the true value and EnKF estimate) produce negligible effects on the tissue temperature at the sensor location, therefore making it difficult to discern the true underlying parameter value from the available observations.
We further note that, while only observing data at the surface sensor location, the filter is able to well estimate the tissue temperature at the two locations tracked below the tissue surface, with wider uncertainty bounds as the location becomes farther from the surface.

\subsection{Identifying Time-Varying Absorption and Scattering}

In the next experiment, we repeat a similar procedure as above, but this time altering the tissue optical properties such that they change over time during laser exposure. This scenario simulates shifts in the tissue optical properties that may occur during the actions of a laser procedure. Laser-induced alterations in the absorption and scattering coefficients have been documented in prior literature~\cite{Vogt2018,Nagarajan2020,Vogel2003}, and our current understanding is that these changes occur due to the tissue's exposure to heat created by the laser.

\begin{figure}[t!]
  \centerline{\includegraphics[width=\linewidth]{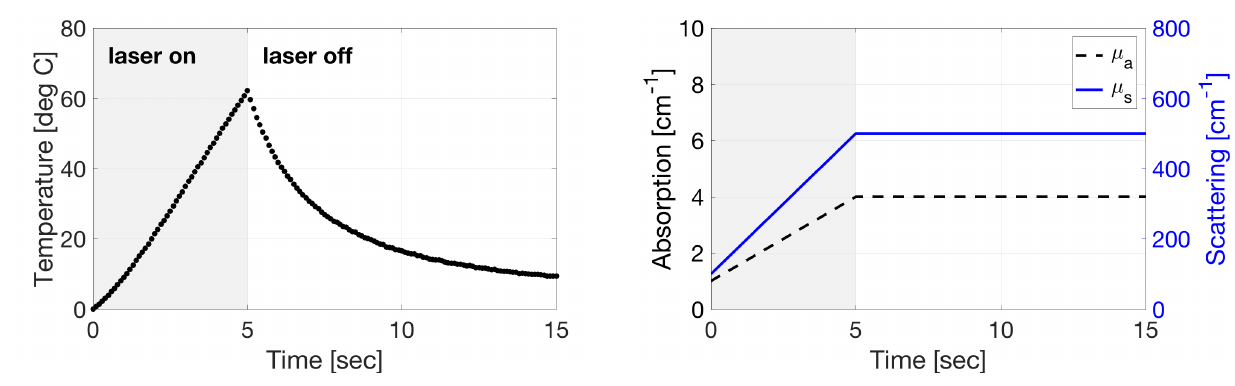}}
  \caption{Sensor data generated with time-varying tissue optical properties. (Left) Surface temperature data observed via virtual sensor at the measurement location shown in Figure~\ref{fig:simulation-setup}. (Right) The corresponding absorption and scattering coefficients, as defined in \eqref{Eq:TV_absorption} and \eqref{Eq:TV_scattering}, respectively. The volumetric heat capacity and thermal conductivity are given in Table~\ref{tab:tissue-parameters}. In each plot, the laser pulse duration is shaded in gray.}
  \label{fig:data_TV}
\end{figure}

To simulate these shifts, we model both the absorption and
scattering coefficients as continuous piecewise functions that
increase linearly during the laser pulse and remain constant when 
the laser is turned off. Recent studies in the biomedical imaging literature indicate that ramp-increasing functions are a valid approach to model the shift in the optical properties of laser-irradiated tissue \cite{Baez2020}.
More specifically, we let
\begin{equation}\label{Eq:TV_absorption}
    \mu_a(t) = \begin{cases} 0.6t+1 & \mbox{if } 0\leq t \leq 5 \\ 
    4 & \mbox{if } 5 < t \leq 15 \end{cases}
\end{equation}
and
\begin{equation}\label{Eq:TV_scattering}
    \mu_s(t) = \begin{cases} 80t+100 & \mbox{if } 0\leq t \leq 5 \\ 
    500 & \mbox{if } 5 < t \leq 15 \end{cases}
\end{equation}
respectively. 
Figure~\ref{fig:data_TV} plots these time-varying absorption and scattering coefficients, along with the simulated sensor observations that correspond.
We note that the EnKF does not assume any knowledge of these 
relations, and the goal of this experiment is
precisely to verify if the filter is able to track the
absorption and scattering coefficients as their values change over time.

\begin{figure}[t!]
  \centerline{\includegraphics[width=\linewidth]{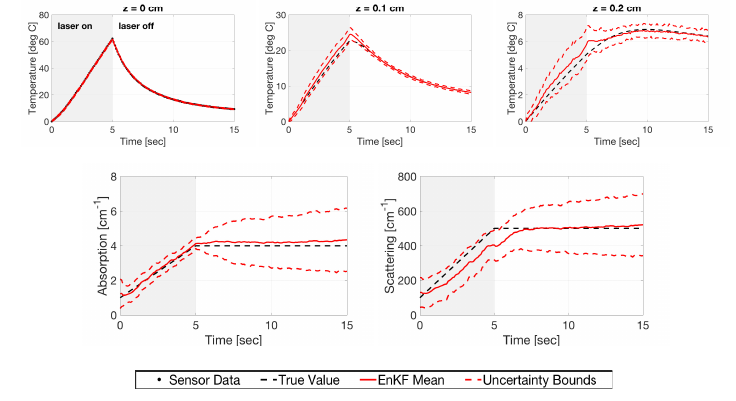}}
  \caption{Time-varying parameter estimation results. (Top) EnKF tissue temperature estimates at three locations on and below the surface of the tissue. (Bottom) EnKF estimates of the absorption and scattering coefficients. In each plot, the laser pulse duration is shaded in gray, the true value of the temperature or parameter is shown in dashed black, the EnKF mean is shown in solid red, and the $\pm 2$ standard deviation uncertainty bounds around the mean are shown in dashed red. The sensor data are shown in black markers on the temperature plot when $z=0$, overlaid with the EnKF estimate and uncertainty bounds.}
  \label{fig:results_TV}
\end{figure}

Figure~\ref{fig:results_TV} displays the resulting estimates of $\mu_a(t)$ and $\mu_s(t)$, along with the corresponding tissue temperature estimates at the three aforementioned locations. These results show that the filter is able to well track the change in absorption throughout the duration of the experiment. The increase in the scattering coefficient is more difficult to track during the laser pulse, but the filter is able to capture its overall behavior and identify the true constant value shortly after the laser is turned off. In both cases, the uncertainty bounds for the time-varying coefficients become increasingly wider once the laser is off. The estimates of the tissue temperature at and below the surface remain accurate throughout the experiment.

\subsection{Estimating Additional Unknown Tissue Properties}

In the previous two experiments, we estimate the unknown absorption and scattering coefficients while assuming that the remaining tissue physical properties (in particular, the volumetric heat capacity, $c_v$, and thermal conductivity, $k$) are known and fixed to their true values. However, since these properties may also be unknown or uncertain in realistic settings, we note that the framework described in Section~\ref{sec:methods} can be extended to accommodate the estimation of additional unknown or uncertain tissue parameters. To do this, we modify the vector of unknown parameters to include $c_v$ and $k$, so that $\theta = (\mu_a, \mu_s, c_v, k)$, and proceed with the estimation as previously described. Note that the additional unknown parameters increase the complexity of the inverse problem by introducing additional uncertainty into the solution of the laser-tissue interaction model at each time step.

\begin{figure}[t!]
  \centerline{\includegraphics[width=\linewidth]{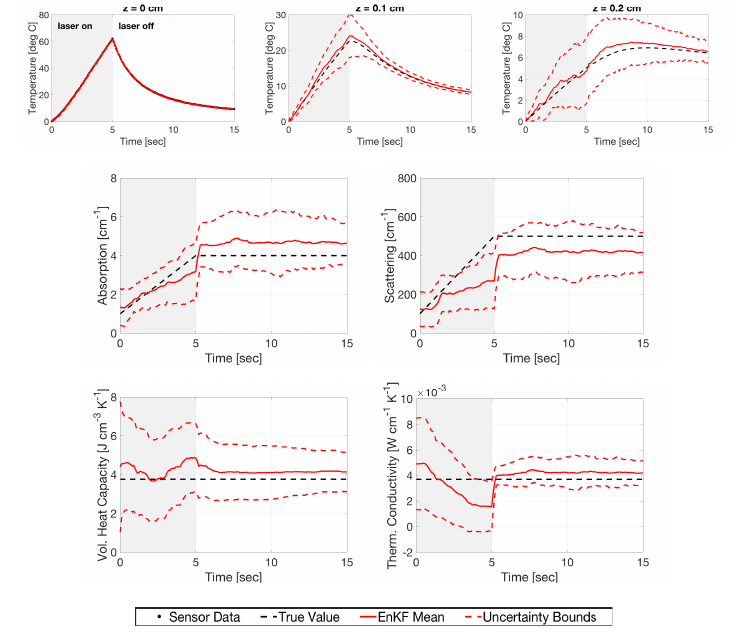}}
  \caption{Additional tissue parameter estimation results. (Top) EnKF tissue temperature estimates at three locations on and below the surface of the tissue. (Bottom) EnKF estimates of the absorption and scattering coefficients, as well as the volumetric heat capacity and thermal conductivity. In each plot, the laser pulse duration is shaded in gray, the true value of the temperature or parameter is shown in dashed black, the EnKF mean is shown in solid red, and the $\pm 2$ standard deviation uncertainty bounds around the mean are shown in dashed red. The sensor data are shown in black markers on the temperature plot when $z=0$, overlaid with the EnKF estimate and uncertainty bounds.}
  \label{fig:results_AllParams}
\end{figure}

In this experiment, we assume the same sensor data as in Figure~\ref{fig:data_TV} and aim to estimate both the time-varying absorption and scattering coefficients as well as the constant volumetric heat capacity and thermal conductivity parameters. Figure~\ref{fig:results_AllParams} shows the resulting EnKF estimates of $\mu_a(t)$, $\mu_s(t)$, $c_v$, and $k$, along with the corresponding tissue temperature estimates at the three specified locations. Despite the additional complexity, results show that all four parameters are generally well tracked, with the true parameter values in each case lying within the EnKF uncertainty bounds. Compared to the results in Figure~\ref{fig:results_TV}, the effects of introducing the additional unknowns are seen in the estimates of $\mu_a(t)$ and $\mu_s(t)$, where the EnKF has more difficulty tracking the true time-varying functions. However, the filter is able to identify the ramp-like increases in both $\mu_a(t)$ and $\mu_s(t)$ during the laser pulse and converge to constant values after the laser is turned off (with relative errors at time $t=15$ on the order of $10^{-1}$ in each case), thereby capturing the overall shape of the underlying functions. The final EnKF estimates of $c_v$ and $k$ at time $t=15$ are slightly higher than the true parameter values, also with relative errors on the order of $10^{-1}$ in each case. The temperature profiles at $z=0.1$ and $z=0.2$ cm below the tissue surface are both well estimated by the EnKF mean but with wider uncertainty bounds than in Figure~\ref{fig:results_TV}, which is reasonable due to the additional modeling uncertainties introduced.

\section{Discussion} 
\label{sec:discussion}

In this work, we propose a novel online procedure for identifying tissue optical properties during thermal laser-tissue interactions. The proposed method utilizes ensemble Kalman filtering and tissue temperature measurements obtained via thermal sensor technology to estimate and track changes in the absorption and scattering coefficients during laser exposure. Through simulated experimental studies, we demonstrate the viability of our approach in identifying both constant and time-varying optical coefficients, as well as to track the tissue temperature at unmeasured locations beneath the tissue surface. 

In the constant coefficient case, where the absorption and scattering coefficients remain constant over time, the results in Figure~\ref{fig:results_constant} show that the filter is able to well identify and track with reasonable accuracy the underlying true coefficient values throughout the pulse duration. We observe a tightening of the uncertainty bounds after the pulse ends, indicating that the filter becomes more certain about the estimates. While the scattering estimate drifts a bit higher after the pulse, the true value remains within the uncertainty bounds of the estimate. As illustrated in Figure~\ref{fig:sens}, the temperature output at the sensor location is less sensitive to the scattering coefficient than the absorption coefficient, and the resulting EnKF estimate (within 25 cm$^{-1}$) is therefore reasonable with respect to the model. 

The time-varying coefficient case presents a challenging problem in which there is a linear increase in both the absorption and scattering coefficients during the laser pulse, which then switch to constants after the pulse. The results in Figure~\ref{fig:results_TV} show that the filter is able to identify and well track the increase in both coefficients during the pulse as well as determine the constant value after the pulse. Once the laser is turned off, the quick transition from linearly increasing to constant scattering without additional laser dynamics results in larger uncertainty over time. We note that the proposed methodology is not limited to linear or ramp-increasing change in the parameters and that other time-varying functions for the parameters could be considered.

Presenting a further challenge, we also demonstrate the capability of the framework in estimating additional unknown tissue physical properties (specifically, the volumetric heat capacity and thermal conductivity) along with the time-varying absorption and scattering coefficients. Despite the additional modeling uncertainties, the results in Figure~\ref{fig:results_AllParams} show promise in the filter's ability to reasonably approximate all four unknown tissue parameters in situations when the volumetric heat capacity and thermal conductivity of the tissue may not be known in advance of a laser action.

One of the main benefits of the method we theorize in this paper is that its implementation simply requires the use of a thermal sensor to monitor the tissue temperature. Advances in thermal sensing technology recently enabled the creation of miniaturized infrared thermal imagers that can be easily integrated in a clinical setup~\cite{Lin2017}. We envision two possible ways to use the proposed identification method for medical laser procedures:
(i) before a procedure, a laser
could ``probe'' the tissue with a short low-power pulse and
identify its optical parameters to inform the planning
of subsequent laser actions;
(ii) the tissue coefficients could be continuously
monitored during the execution of a laser action, and this
information could be used to dynamically change the laser inputs to adapt to changes in the tissue optical properties or to signify a stopping point for the action.
We note that this latter option is made possible by the fact that the EnKF applied in this paper is capable of estimating time-varying parameters. Similar implementations of the EnKF have recently been utilized for tracking time-varying parameters in biological applications~\cite{Arnold2019,Campbell2020}.

Another benefit of this work is that it implicitly provides a method to monitor the \textit{internal} tissue temperature during laser irradiation, where this would normally require the use of an invasive sensor (e.g., a thermocouple deployed into the tissue).
Being able to monitor the tissue temperature 
is a long-standing problem in 
laser surgery~\cite{Pardo2014,Pardo2015}, as this capability
is vital to anticipate and prevent the onset of thermal injuries.
The results in Figures~\ref{fig:results_constant}, \ref{fig:results_TV}, and \ref{fig:results_AllParams} show the 
EnKF successfully tracking the tissue temperature
at locations below the surface ($z=$ 0.1 cm, 0.2 cm).
It is important to remark that the filter did not 
receive any temperature data from these 
locations, as in our study we assumed the use of a
sensor that only provided superficial measurements at a single surface location.
Tracking of the internal tissue temperature
is enabled by the fact the EnKF was augmented to
include the tissue temperature, as in \eqref{eq:ensemble}, therefore
enabling the prediction of temperature
dynamics across the entire volume.

The evidence reported in this paper was generated in simulation, assuming tissue with uniform properties, to validate and provide proof-of-principle of the
proposed estimation method.
In future work, we aim to further corroborate the viability of the method with
real laser-tissue interaction experiments.
In moving towards a more realistic implementation, we also aim to address possible limitations due to the computational complexity of the proposed
identification method.
Although the filter itself is computationally inexpensive, 
the thermal model used in the prediction step 
(Section~\ref{sec:prediction-step}) relies on a Monte Carlo
method to simulate light absorption.
The use of Monte Carlo simulation may create a scalability
barrier, especially if one wishes to use the method described
in this paper to monitor the tissue optical properties 
online during a medical procedure.
We plan to investigate the use of alternative
thermal models that offer a different balance between the
accuracy of the temperature predictions and computational
complexity.
We also plan to study the capability of the proposed approach in tracking changes in the tissue optical properties over a series of laser pulses.

\section{Conclusion}
\label{sec:conclusion}

This paper introduces a method to identify the optical properties of tissue (namely, the absorption and scattering coefficients) using ensemble Kalman filtering and tissue temperature measurements obtained via thermal sensor.
Knowledge of these coefficients is vital to enable accurate modeling and control of the laser-tissue interactions during laser procedures.
The contributions made in this work have the potential to enable real-time detection of changes in tissue properties during laser surgery, where this information is key in the planning and execution of subsequent laser actions.
Through simulated experiments of increasing complexity, we demonstrate the viability of the proposed approach and discuss aspects to be addressed in future work.



\section*{Acknowledgments}

This work was partially supported by the National Science Foundation under grant number NSF/DMS-1819203 (A. Arnold). The authors contributed equally to this work.


\section*{ORCID iDs}
Andrea Arnold: \url{https://orcid.org/0000-0003-3003-882X}

\vspace{5pt}

\noindent Loris Fichera: \url{https://orcid.org/0000-0001-7347-9479}



\bibliography{laser_refs}{}

\end{document}